\documentclass[prb,nofootinbib,amsfonts,tightenlines,twocolumn,superscriptaddress]{revtex4-1}
\usepackage{graphicx}
\usepackage{dcolumn}
\usepackage{bm}
\usepackage{caption}
\usepackage{subcaption}
\usepackage{mathrsfs}
\usepackage{dsfont}
\usepackage{physics}
\usepackage{natbib}
\usepackage{xcolor}
\usepackage{wrapfig}
\usepackage{upgreek}
\usepackage[utf8]{inputenc}
\usepackage{xpatch}
\usepackage{amsmath}
\usepackage{amssymb}

\usepackage[normalem]{ulem}

\begin{document}

\title{Anomalous topology and synthetic flat band in multi-terminal Josephson Junctions}
\author{Aabir Mukhopadhyay} 
\email{aabir.riku@gmail.com, \\ ORCID ID : 0000-0001-6465-2727}
\affiliation{Indian Institute of Science Education \& Research Kolkata,
Mohanpur, Nadia - 741 246, 
West Bengal, India}

\author{Udit Khanna} 
\email{udit.khanna.10@gmail.com, \\ ORCID ID : 0000-0002-3664-4305}
\affiliation{Department of Physics, Bar-Ilan University, Ramat Gan 52900, Israel}

\author{Sourin Das}
\email{sourin@iiserkol.ac.in, \\ ORCID ID : 0000-0002-8511-5709}
\affiliation{Indian Institute of Science Education \& Research Kolkata,
Mohanpur, Nadia - 741 246, 
West Bengal, India}
\date{\today}

\begin{abstract}
Andreev bound states trapped in a multi-terminal Josephson junction (JJ) can be assigned a synthetic band topology owing to their periodic dependence on the Josephson phase bias. We demonstrate that the BdG symmetry adds a twist to this  topological character, i.e,  gap closing points  may or \textit{may not} correspond to change of Chern number, hence extending the standard paradigm for topological bands. We further show that the topology of Andreev bands depends only on the scattering matrix of the junction and is independent of the topological nature of superconductors forming the JJ hence indicating a universal behaviour of multi-terminal JJ. We also show that the chiral junction, supported by quantum Hall state at the junction region, leads to flat Andreev bands (implying absence of DC Josephson effects) that are devoid of Berry curvature ( implying absence of AC Josephson effects). Such electrically inert JJ may be useful for storage of quantum information in future quantum devices.
\end{abstract}

\maketitle
\textit{Introduction:}~In the realm of quantum condensed matter physics, the application of the mathematical concept of topology in the context of insulators and superconductors, has garnered considerable attention. The topological nature of these systems is intrinsically linked to the presence of a gap in their energy spectrum. This gap acts as a protective barrier, signifying that altering the system’s energy landscape through gap closure can lead to phase transitions, causing change in the system’s topological quantum number. In an insightful work~\cite{ezawa_SR_3_174504}, Nagaosa highlighted the possibility of a topological transition without closing the gap by varying the system's symmetry across the transition. Certain gapless systems, such as Weyl semimetals, may also possess topological character~\cite{shuichi_NJP_9_356, wan_PRB_83_205101, burkov_PRL_107_127205, burkov_PRB_84_235126, zyuzin_PRB_85_165110, hosur_PRL_108_046602, xu_sc_349_613, xu_NP_11_748, lv_PRX_5_031013, lv_NP_11_724, lu_sc_349_622, jia_NM_15_1140}. In this Letter, we demonstrate a distinct scenario where, in contrast to the previous cases, the topology of a gapped system may or \textit{may not} change at a gap closing (without any change in the system's symmetries). 
This intriguing anomaly, which is the central theme of our study, raises questions about the interplay between gap closure and topology change.

We focus on subgap states of Josephson junctions (JJ) i.e. the Andreev bound states (ABS), the spectrum of which is necessarily periodic in the $n$ superconducting phases ($\phi_{i}$ for $i = 1, \dots, n$) of an $n$-terminal JJ. Hence, each ABS forms a synthetic band in the $n-1$-dimensional space spanned by $\{\phi_{2}, \phi_{3}, \dots, \phi_{n}\}$ (defined with respect to $\phi_1$) that may be characterized with topological quantum numbers, as is done for lattice Hamiltonians~\cite{ja2016short, kaufmann_RMP_28_1630003, topoCondMat, read_the_docs}. A typical example of such topological indices is the Chern number which may be defined for even dimensional bands. Here, we show that the topology of the Andreev bands (characterized through the Chern number) is decoupled from the topological nature of the superconductors forming the junction and instead depends only on the scattering matrix ($S$) describing the junction, though they share an interesting complementarity in terms of the Berry curvature and Chern number.


We also demonstrate that is it is possible to host perfectly flat Andreev bands in the minimal but non-trivial case of  three-terminal Josephson junction (3-JJ) when a chiral $S$-matrix is imposed at the junction; a scenario which may be motivated by allowing for a $\nu=1$ quantum Hall state or a anomalous Hall state at the junction~\cite{oshikawa_JSM_2_02008,khanna_PRB_105_L161101,vignaud2023evidence}. We find that such flat bands are devoid of any electrical response (no DC or AC Josephson effect) and hence may find application as a safe storage unit of quantum information.   

\begin{figure}[t]
	\centering
	\includegraphics[width=1.0\columnwidth]{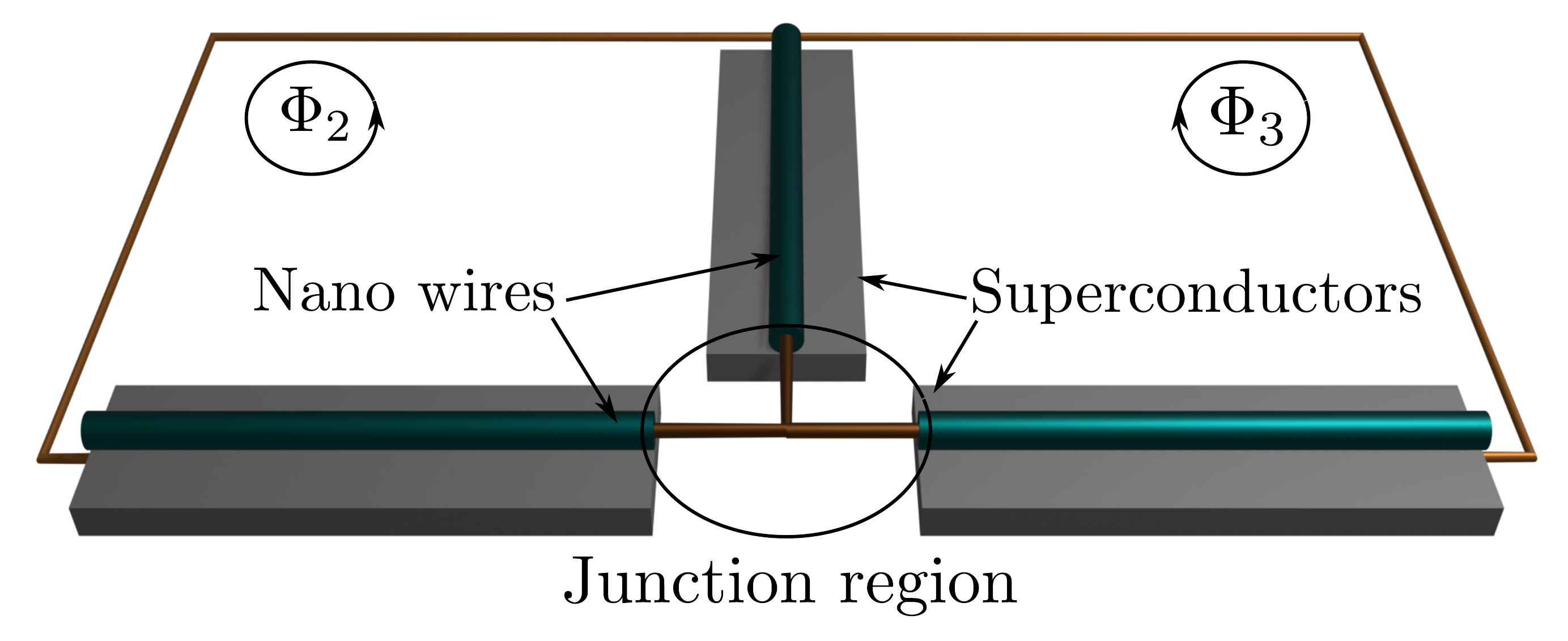}
	\caption{Schematic diagram of a three-terminal Josephson junction realized with semiconducting nano-wires and proximity induced superconductivity. The independent phase-biases $\phi_2$ and $\phi_3$ (measured with respect of $\phi_1$) can be tuned through two independent flux through the circuit $\Phi_2=(\hbar \phi_2)/(2e)$, $\Phi_3=(\hbar \phi_3)/(2e)$.}
	\label{SetUp}
\end{figure}

\begin{figure*}[htbp]
	\centering
	\includegraphics[width=1.9\columnwidth]{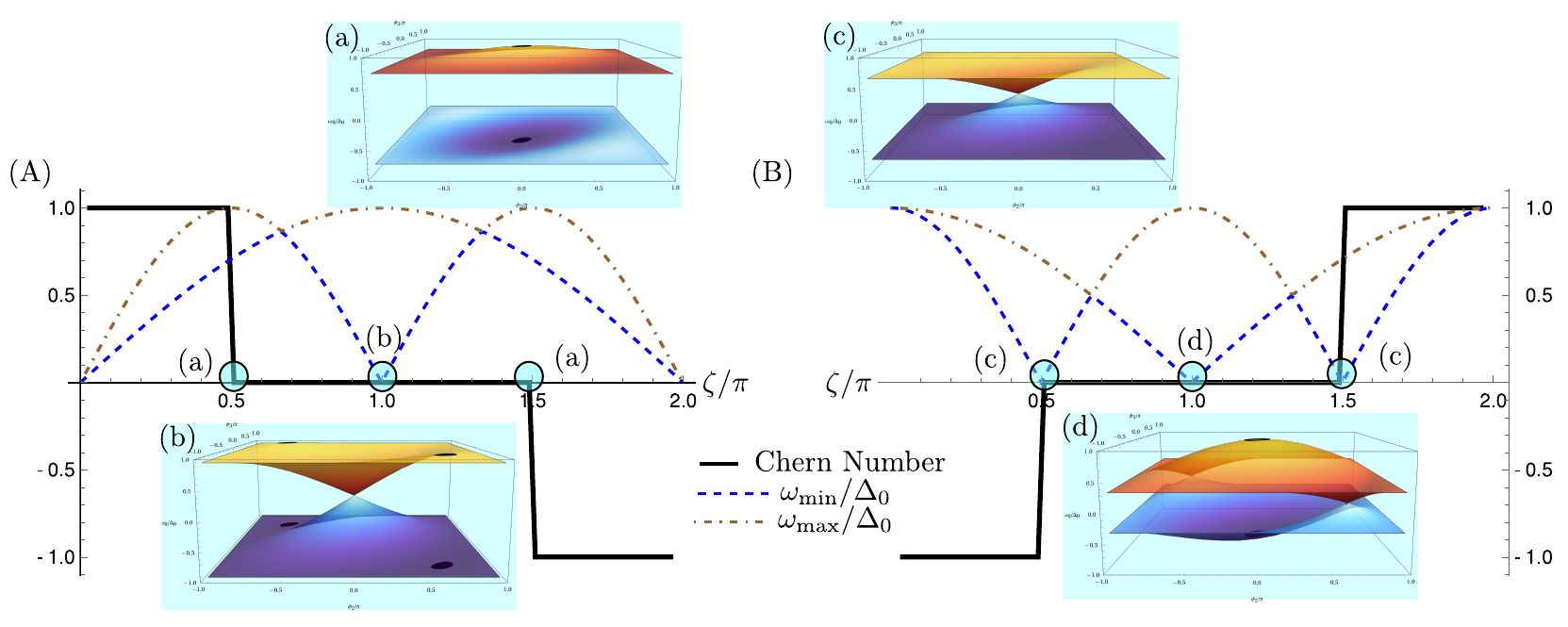}
	\caption{Variation of the Chern number of the negative energy ABS band ($\omega_{-}$), the minimum ($\omega_{\text{min}}$) and maximum energy gap ($\omega_{\text{max}}$) in the Andreev spectrum over the family of S-matrices defined in Eq.~\ref{Gell_Mann_Scat_1}. The superconductors forming the 3-JJ may be (A) topological ($\eta = -1$) or (B) non-topological ($\eta = +1$). The insets (a-d) depict the dispersion of the non-trivial bands ($\omega_{\pm}$) at parameters for which (a,c) the Chern number changes or (b,d) the band gap closes. The black shaded regions are the points where the bands touch the continuum.}
	\label{MaxMinGapChern}
\end{figure*}

\textit{System and the Chern number:}~The system we are interested in is a 3-JJ (Fig.~\ref{SetUp}), based on quadratic dispersion of 1-D electron gas. Such junctions can be realized effectively with semiconducting nano-wires and conventional $s$-wave superconductors by means of proximity effect\cite{lutchyn_PRL_105_077001, oreg_PRL_105_177002, shim_NC_5_4225, larsen_PRL_115_127001, lange_PRL_115_127002, krogstrup_NM_14_400, chang_NN_10_232}. The junction region can be described in terms of a $3 \times 3$ scattering matrix $(S)$. Throughout this work, we assume that the length of the junction is much smaller than the coherence length of the superconductors and that this $S$-matrix is independent of energy.

For energies below the superconducting gap $\Delta$ (taken to be equal in all terminals for simplicity), the Andreev reflection amplitude at a normal metal-superconductor junction is unimoduler and repeated Andreev reflections results in the formation of ABS at certain discrete energy levels ($\omega$). The Andreev spectrum may be found through the following eigenvalue equation, which relies on the wavefunction ($\psi$) being unique after two consecutive Andreev reflections~\cite{beenakker_PRL_67_3836, deb_PRB_97_174518, self_PRB_103_144502}, 
\begin{equation}
	S e^{i \varphi} S^* e^{-i \varphi} \psi_i = \eta e^{2 i \theta} \psi_i .
	\label{ABS_equation}
\end{equation}
Here, $\theta=\arccos(\omega/\Delta)$ and $\eta=\pm 1$ labels the nature of the superconductors: $+1$ ($-1$) for $s$-wave ($p$-wave) pairing. The matrix $S (S^*) \in U(3)$~\footnote{For quadratic dispersion, the charge conjugation operator is $\mathcal{C}=\mathcal{K}$, where $\mathcal{K}$ is the operation of complex conjugation.} describes the scattering of electrons (holes) at the junction region and $\varphi$ is a diagonal matrix with diagonal elements $\{\phi_1=0,\phi_2,\phi_3\}$. Equation (\ref{ABS_equation}) has 3 solutions for the ABS energy, one of which (labelled $\omega_{0}$) is always independent of \{$\phi_2,\phi_3$\} (a trivial flat band), while the other two (denoted as $\omega_\pm$) are non-trivial with $\omega_+=-\omega_-$ due to the BdG symmetry. This holds for both values of $\eta$.
The Berry curvature ($B_{i,23}$) and Chern number ($\text{Ch}_{i,23}$) for these bands may be defined in terms of $\psi_i$ as,
\begin{equation}
	B_{i,23}(\phi_2,\phi_3)=-2 \text{Im}\left[\dfrac{\partial \psi_i^{\dagger}}{\partial \phi_2} \dfrac{\partial \psi_i}{\partial \phi_3}\right],
\end{equation}
\begin{equation}
	\text{Ch}_{i,23}=\dfrac{1}{2 \pi} \int_{0}^{2 \pi} \int_{0}^{2 \pi} d\phi_2 d\phi_3 B_{i,23} .
\end{equation}
The standard theory of topological bands predicts that the Chern number of $\omega_{\pm}$ bands would change when the gap ($\delta \omega=\text{Min}[\omega_+]$) closes ($\delta \omega=0$~\cite{meyer_PRB_103_174504}. 
The finite superconducting gap allows for the possibility of the ABS bands touching the continuum states, i.e., $\text{Max}[\omega_+]= \Delta$. In this case, the Chern number for the $\omega_{\pm}$ bands becomes ill-defined and may also lead to a change of the Chern number~\cite{meyer_PRL_119_136807}. In what follows, we will show that there exists a scenario beyond this conventional wisdom. 

To demonstrate this, we consider a 3-JJ that may be described by a one-parameter ($\zeta$) family of real $S$-matrices, defined as, 
\begin{equation}
	S=\exp(i \dfrac{\zeta}{\sqrt{3}} (\lambda_2 + \lambda_5 +\lambda_7)). 
	\label{Gell_Mann_Scat_1}
\end{equation}
Here $\lambda_2, \lambda_5$ and $\lambda_7$ are the three purely imaginary Gell-Mann matrices (see Supplemental Material~\ref{Gell_Mann_Matrices}).
Figure~\ref{MaxMinGapChern} shows the maximum and minimum values of the gap ($\delta \omega$) over the entire band ($\phi_2-\phi_3$ plane) and the Chern number of the $\omega_-$ band for all $\zeta$. 


Two striking facts emerge from Fig.~\ref{MaxMinGapChern}: \textit{(i)} it demonstrates that the case of topological superconductors ($\eta=-1$) and the non-topological case ($\eta= 1$) are related via a complementarity. Note that the Chern numbers have opposite sign for the same value of $\zeta$ (same $S$ matrix), and the gap closing point ($\delta \omega=0$) in one case is replaced by the ABS band touching the continuum ($\text{Max}[\omega_+]= \Delta$) in the other and vice versa. \textit{(ii)} The Chern number does not change at $\zeta=\pi$ even though there is an odd number (3) of singular points in the band structure: for $\eta = -1$, the Andreev bands touch each other at $\{\phi_2, \phi_3\}=\{0,0\}$ and they touch the continuum at two distinct points, symmetrically placed about $\{0,0\}$, along the line $\phi_2=-\phi_3$. The case for $\eta=1$ is related to this via the complementarity described above. These synthetic bands present a previously unexplored anomalous situation where a gap closing is necessary but not a sufficient condition for the change of the Chern number.




\textit{Formalism:} To develop an understanding of this anomalous behaviour we start with a formulation which treats the case of non-topological and topological superconductor on an equal footing. We start by noting  that as far as Berry curvature and Chern number are concerned, they are connected to the eigenstates of Eq.~\ref{ABS_equation}, and the only source of difference between topological and non-topological 3-JJ is an overall phase of $\pi$. Thus, both kind of junctions can be described in terms of the eigensystem of the matrix:
\begin{equation}
	Q=\Sigma \Sigma^*= S e^{i \varphi} S^* e^{-i \varphi}
	\label{Q_and_Sigma_matrices}
\end{equation}
where $\Sigma=S e^{i \varphi}$. Since $S, e^{i \varphi} \in U(3)$, so is $Q$. The eigenvalues of the matrix $Q$ take the form $\{e^{i \lambda},1,e^{-i \lambda}\}$ (See Supplemental Material \ref{Eigenvalues_of_Q}) 
These eigenvalues of $Q$ can be equated to $\eta e^{2 i \theta}$ to determine the ABS energies. Note that, the eigenvalue 1 (corresponding to the trivial flat band), when equated to $\eta e^{2 i \theta}$ gives $\omega_0=0$ for $\eta=-1$ and $|\omega_0|=\Delta$ for $\eta=1$. We shall denote the corresponding eigenfunctions of Q as $\{ \psi_{+}, \psi_0, \psi_{-}\}$. The Chern number of the trivial ABS band (corresponding to the eigenvalue 1 in Eq. \ref{Eigen_value_equation_SM}) is always zero, and the Chern numbers of the non-trivial bands (corresponding to the eigenvalues $e^{i \lambda}$ and $e^{-i \lambda}$ in Eq. \ref{Eigen_value_equation_SM}) come with equal magnitude and opposite sign, due to particle-hole symmetry.

Below we identify an effective Hamiltonian for the junction   
given by
\begin{align}
	\mathcal{H}\psi_i =\dfrac{1}{2 i}(Q-Q^{\dagger}) \psi_i &=\left\{ \frac{e^{i \lambda}-e^{-i \lambda}}{2 i},0 ,\frac{e^{-i \lambda}-e^{i \lambda}}{2 i}\right\} \psi_i	\nonumber \\
	&=\{\sin(\lambda),0,-\sin(\lambda)\} \psi_i	\nonumber\\
	&= \eta \left\{ +\Omega,0 ,-\Omega \right\}\psi_i
 \label{Effective_H}
\end{align}
such that it has the ABS wavefunctions $\{ \psi_{+}, \psi_0, \psi_{-}\}$ as its eigenfunction owing to the fact that  $[\mathcal{H},Q]=0$ and the corresponding eigenvalues are give by $\Omega=(2\omega \sqrt{\Delta^2-\omega^2})/\Delta^2$. This equation also makes it apparent that $\eta=1 \rightarrow \eta=-1$ implies $\{\psi_+,\psi_0,\psi_-\}\rightarrow \{\psi_-,\psi_0,\psi_+\}$. Hence the local Berry curvature in the $\{\phi_1,\phi_2\}$-plane and the Chern number of ABS bands follow complementarity between the $\eta=1$ and the  $\eta=-1$ case.  


Identifying $\mathcal{H}$ as an effective junction  Hamiltonian offers  advantages over directly studying eigenvalue Eq. \ref{ABS_equation} as far as Chern number of Andreev Bands are concerned: \textit{(i)} It provides  an unified framework for both non-topological and topological 3-JJ. \textit{(ii)} It keep the essence of BdG symmetry intact while mapping both types of band touching points \--- gap closure and ABS bands touching continuum \--- to the gap-closing point so that it mimics a standard topological insulator Hamiltonian where gap closing point are the only special points in the parameter space and it helps in understanding  the anomalous topology in a geometric fashion as we will see now.


$\mathcal{H}$ can be expanded in terms of 8 Gell-Mann matrices (See Supplemental Material \ref{Gell_Mann_Matrices}) as :
\begin{equation}
    \mathcal{H}=\vec{a}.\tilde{\lambda}=\sum_{i=1}^{8} a_i \tilde{\lambda}_i.
\end{equation}
Note that one can choose three $SU(2)$ sub-algebras: $u_1=\{\tilde{\lambda}_1,\tilde{\lambda}_2,\tilde{\lambda}_3\}$, $u_2=\{\tilde{\lambda}_4,\tilde{\lambda}_5,x\}$ and $u_3=\{\tilde{\lambda}_6,\tilde{\lambda}_7,y\}$; such that $x=(\tilde{\lambda}_8+\tilde{\lambda}_3)/2$ and $y=(\tilde{\lambda}_8-\tilde{\lambda}_3)/2$. Projection of $\mathcal{H}$ can be taken into each of these sub-algebra spaces, which we call $\mathcal{H}_1$, $\mathcal{H}_2$, and $\mathcal{H}_3$ respectively. Due to the $SU(2)$ nature of these sub-algebras, these projected Hamiltonians have the form $\mathcal{H}_i=\Vec{B_i}.\vec{\sigma}$ (where $\sigma$ are the natural generalization of Pauli matrices to $SU(3)$) and share similar spectral structure as that of $\mathcal{H}$ i.e. $\{ +|B_i|^2 ,0, -|B_i|^2\}$. It turns out that the gap closing in $\mathcal{H}$ and in its projected Hamiltonians $\mathcal{H}_1$, $\mathcal{H}_2$, $\mathcal{H}_3$ always occurs simultaneously in the parameter space of $\phi_2-\phi_3$ (See Supplemental Material \ref{Effective_Hamiltonian}). Now from conventional wisdom of gap-protected topology, we know that the change of topology is tied to the gap-closing of the spectrum. Thus, rather than visualizing the gap-closing and hence the change in topology of the 8 dimensional parameter space of $\mathcal{H}$ which is non-trivial, it is straight forward to geometrically visualize the gap-closing and the change in topology of each of the projected Hamiltonians, which resides on the parameter space of 3 dimensional vector space $\vec{B}_i$, by investigating how many times does the torus formed by the image of Brillouin zone ($0<\phi_2,\phi_3\leq 2\pi$) in any of these projected spaces contail the origin. We have checked over 500 random $S$ matrices, chosen from Circular Unitary Matrix distribution, and observe that the Chern numbers of the positive (negative) bands of $\mathcal{H}$, $\mathcal{H}_1$, $\mathcal{H}_2$ and $\mathcal{H}_3$ always coincides which implies not only the gap-closing point but also the topology itself of $\mathcal{H}$ and its projected Hamiltonians coincide. Now this formulation will be used to explain the origin of anomalous behaviour discussed earlier.

\begin{figure}[t]
	\centering
	\includegraphics[width=\columnwidth]{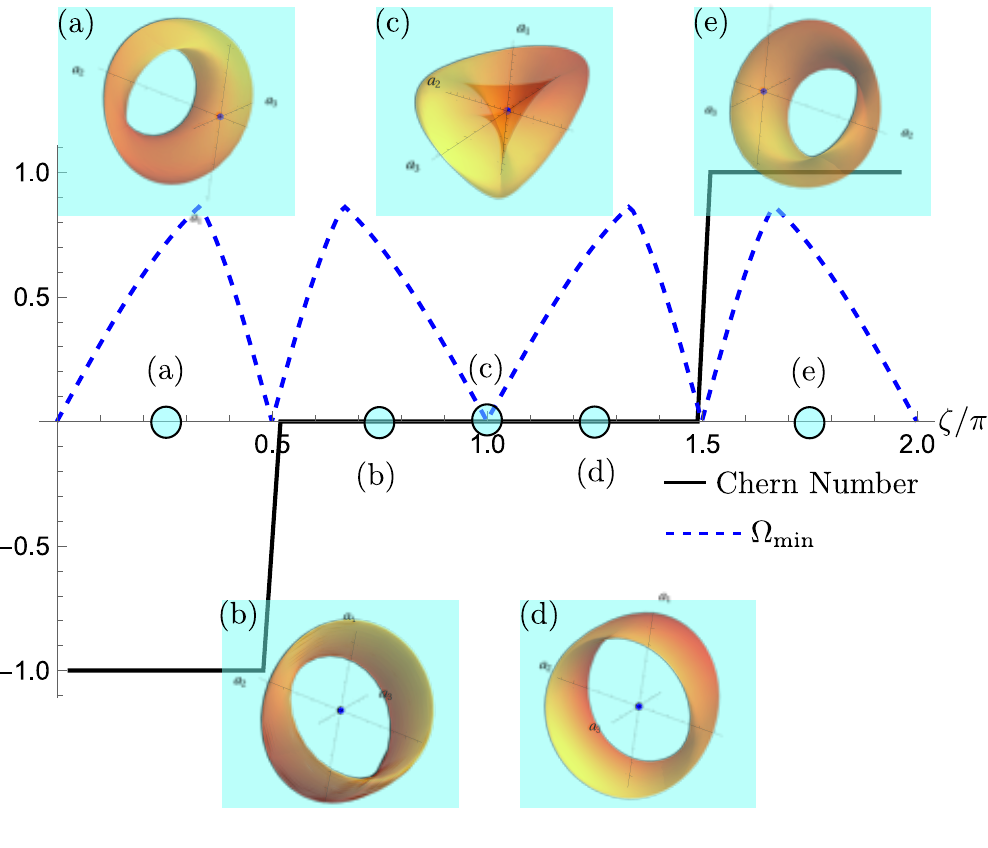}
	\caption{Chern number corresponding to wave functions $\psi_{-}$ and the gap in the spectrum of $\mathcal{H}$. Torus formed by the image of Brillouin zone in the projected Hamiltonian $H_1$ are also shown at some critical points.}
	\label{MinGapChernContour}
\end{figure}

\textit{Gap-closing without topology change:} Referring back to the real scattering matrix given in Eq. \ref{Gell_Mann_Scat_1}, we construct the corresponding $\mathcal{H}$ and its projections. To obtain a geometric view of the topological transitions, we study the torus formed by the parametric plot (as a function of $\phi_2$, $\phi_3$) of $\vec{B}_i$ corresponding to the projected Hamiltonian $\mathcal{H}_i$ along the parameter $0 \leq \zeta \leq 2\pi$ and check if the singular point corresponding to $\Vec{B}_i=\Vec{0}$ lies inside or outside the torus. Being inside will correspond to a finite Chern number while being outside will correspond to Chern number being zero. 

We observe that within the parameter range of $0 \leq \zeta \leq \pi/2$ and $3 \pi/2 \leq \zeta \leq 2\pi$, the singular point resides inside the torus, indicating non-zero Chern number; whereas within the parameter range $\pi/2 <\zeta < 3\pi/2$, the singular point resides outside the torus, giving rise to zero Chern number. However, at $\zeta=\pi$, the torus touches the singular point, and thereby causing gap-closure in the spectrum, and again bounces back keeping the singular point outside the torus for $\zeta=\pi+\delta$ ($\delta \leq \pi/2$). This is the case for all the projected Hamiltonians $\mathcal{H}_i$ ($i\in \{1,2,3\}$). Thus, in the parameter space, on the both sides this gap-closing point at $\zeta=\pi$, the Chern number corresponding to the projected Hamiltonians does not change. This fact provides a clear understanding of the anomalous behaviour of $\mathcal{H}$ at $\zeta=\pi$ owing to the one-to-one correspondence of gap-closing point and Chern number between $\mathcal{H}$ and its $SU(2)$ projections.



\begin{figure}[t]
	\centering
	\includegraphics[width=\columnwidth]{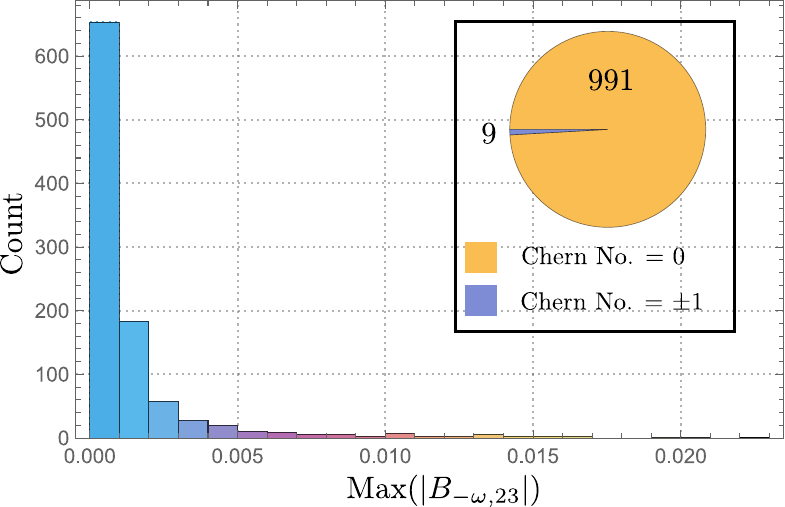}
	\caption{Distribution of the Chern number and maximum of the absolute value of Berry curvature ($B_{-\omega, 23}$) over $\phi_2-\phi_3$ space of the negative energy Andreev band for 1000 random scattering matrices in the neighbourhood of the chiral fixed point.}
	\label{Chern_Berry_1000}
\end{figure}

\textit{Chiral junction:} An interesting point in the parameter space of the S-matrices defined in Eq.~\ref{Gell_Mann_Scat_1} corresponds to  $x=2\pi/3, 4\pi/3$, where the S-matrix is chiral i.e, 
\begin{align}
  S_{\text{chiral}} = \left( \begin{array}{ccc}
    0 & 0 & 1 \\
    -1 & 0 & 0 \\
    0 & -1 & 0
  \end{array}
  \right), \left( \begin{array}{ccc}
    0 & -1 & 0 \\
    0 & 0 & -1 \\
    1 & 0 & 0
  \end{array}
  \right)
\end{align}
Such a junction can be realized by forming a 3-JJ with the edge mode of a $\nu = 1$ (anomalous) quantum Hall state~\cite{oshikawa_JSM_2_02008, khanna_PRB_105_L161101}. Two-terminal Josephson junctions based on quantum Hall edges in graphene nanoribbons were realized recently~\cite{vignaud2023evidence}. Such designs have the advantage that all segments of the edge can be controlled electrically through three metallic gates, forming a 3-way quantum point contact (see fig.~\ref{QHE_QI_Store} for the Supplemental Material for an example). The S-matrix of the junction may be manipulated from chiral to other forms through suitable gate voltages. This offers a way to realize adiabatic manipulations of the Majorana modes residing at the junction (for topological superconductors)~\cite{nagae2023multilocational,pandey2023majorana}. If the system can be adiabatically restored to a chiral form at the end of such operations then it may be used to securely store the result of these manipulations, as we describe below. 


The striking feature of a chiral 3-JJ ($S_{\text{chiral}}$) is that the ABS spectrum becomes independent of the superconducting phases ($\phi_2$, $\phi_3$) which essentially implies a perfectly flat band. As the Josephson current is proportional to the phase derivatives of ABS energy, this implies such junction will have zero DC Josephson effect. The AC Josephson effect has two contributions: one arises from the phase variation of the ABS energy and second is proportional to the Berry curvature of the Andreev bands~\cite{riwar_NC_7_11167, deb_PRB_97_174518}. The first is zero for flat bands, similar to the DC effect. It is straightforward to show that the local Berry curvature corresponding to these flat bands is identically zero over the Brillouin zone ($0\leq \phi_2, \phi_3 \leq 2\pi$), leading to a net zero AC Josephson response. Therefore, such junctions would not only be electrically inert, but would also be robust against small fluctuations of voltage (corresponding to an AC Josephson effect) and magnetic flux / phase (corresponding to the DC effect).

Designing a perfectly chiral 3-JJ can be experimentally challenging. However, a nearly chiral 3-JJ can also be treated as an inert junction for all practical purposes. To demonstrate this point, we carry out a numerical study, over 1000 random scattering matrices which are in the neighbourhood of the chiral junction. To do this, we have considered a scattering matrix of the form

\begin{equation}
    S=\exp\left(\dfrac{i}{\sqrt{3}} \sum_{i=1}^{8} l_i \lambda_i\right)
\end{equation}
where $l_i$ are chosen from Gaussian distribution with mean values $l_2=l_5=l_7=\{ 2\pi/3 , 4\pi/3\}$, $l_1=l_3=l_4=l_6=l_8=0$ and with standard deviation $\pi/10$. Figure~\ref{Chern_Berry_1000} depicts the distribution of the Chern number and the maximum Berry curvature of the positive energy ABS for this ensemble of S-matrices. Since the Chern number is zero in 991 instances (out of 1000), the system is expected to be relatively inert, with a vanishing transconductance in a finite region around the chiral point. Moreover, the fact that the Berry curvature is vanishingly small throughout this region implies that the system would respond very weakly to small fluctuations of the voltage. This establishes the practicality of such junctions.

\textit{Acknowledgement:} AM acknowledges Dr. Tousik Samui for helpful discussions and Ministry of Education, India and IISER Kolkata for funding.
\bibliography{Paper.bib}

\newpage

\onecolumngrid

\newpage

\onecolumngrid

\appendix
\renewcommand\appendixname{Supplemental Material}

\section{Gell-Mann Matrices}
\label{Gell_Mann_Matrices}
The Gell-Mann matrices are 8 linearly independent $3\times 3$ traceless Hermitian matrices which span Lie algebra of the $SU(3)$ group. These matrices are the following:
\begin{equation*}
    \lambda_1 =
    \begin{bmatrix}
        0   &1  &0  \\
        1   &0  &0  \\
        0   &0  &0
    \end{bmatrix}, \hspace{15pt}
    \lambda_2 =
    \begin{bmatrix}
        0   &-i &0  \\
        i   &0  &0  \\
        0   &0  &0
    \end{bmatrix}, \hspace{15pt}
    \lambda_3 =
    \begin{bmatrix}
        1   &0  &0  \\
        0   &-1 &0  \\
        0   &0  &0
    \end{bmatrix},
    \lambda_4 =
    \begin{bmatrix}
        0   &0  &1  \\
        0   &0  &0  \\
        1   &0  &0
    \end{bmatrix}, \hspace{15pt}
    \lambda_5 =
    \begin{bmatrix}
        0   &0  &-i  \\
        0   &0  &0  \\
        i   &0  &0
    \end{bmatrix},
\end{equation*}
\begin{equation*}
    \lambda_6 =
    \begin{bmatrix}
        0   &0  &0  \\
        0   &0  &1  \\
        0   &1  &0
    \end{bmatrix}, \hspace{15pt}
    \lambda_7 =
    \begin{bmatrix}
        0   &0  &0  \\
        0   &0  &-i  \\
        0   &i  &0
    \end{bmatrix}, \hspace{15pt}
    \lambda_8 = \dfrac{1}{\sqrt{3}}
    \begin{bmatrix}
        1   &0  &0  \\
        0   &1  &0  \\
        0   &0  &-2
    \end{bmatrix}
\end{equation*}
For convenience, we have redefined $\lambda_i \rightarrow \Tilde{\lambda}_i$, such that $\Tilde{\lambda}_8 = \sqrt{3} \lambda_8$. Note that,
\begin{equation}
    \Tilde{\lambda}_8=x+y=
    \begin{bmatrix}
        1   &0  &0  \\
        0   &0  &0  \\
        0   &0  &-1
    \end{bmatrix} +
    \begin{bmatrix}
        0   &0  &0  \\
        0   &1  &0   \\
        0   &0  &-1
    \end{bmatrix}
\end{equation}
\begin{equation}
    \Tilde{\lambda}_3=x-y
\end{equation}
Now, one can define three significant $SU(2)$ algebras in terms of the Gell-Mann matrices:
\begin{equation}
    \begin{cases}
       L_1= \{ \Tilde{\lambda}_1, \Tilde{\lambda}_2, \Tilde{\lambda}_3 \}\\
       L_2= \{ \Tilde{\lambda}_4, \Tilde{\lambda}_5, x \}\\
       L_3= \{ \Tilde{\lambda}_6, \Tilde{\lambda}_7, y \}
    \end{cases}
\end{equation}
These sub-algebras are effectively $\{ \sigma_x, \sigma_y, \sigma_z \}$, where $\sigma_i$ are the Pauli's $\sigma$ matrices.

\section{Eigenvalues of the matrix $Q$}
\label{Eigenvalues_of_Q}
The matrix $Q$ has a special form in terms of eigenvalues.
\begin{align*}
	&\text{det.}[Q-\mathbb{I}]	\\
	=&\text{det.}[\Sigma \Sigma^* -\Sigma \Sigma^{\dagger}]	\\
	=&\text{det.}[\Sigma] \text{det.}[\Sigma^*-\Sigma^{\dagger}]	\\
	=&0,
\end{align*}
as $\Sigma^*-\Sigma^{\dagger}$ is an anti-symmetric matrix of order 3 (odd). This implies, there exists a $\psi_0$, such that,
\begin{equation}
	Q \psi_0 = \psi_0
\end{equation}
which means $\Lambda_0=1$ is always an eigenvalue of $Q$. This eigenvalue (when equated to $\eta e^{2 i \theta}$) corresponds to $\omega_0=0$ energy for a topological 3-JJ and $|\omega_0|=\Delta_0$ for non-topological 3-JJ. Let the other two eigenvalues of $Q$ are $\Lambda_{+}$ and $\Lambda_{-}$, then,
\begin{align*}
	&\text{det.}[Q]=\prod \Lambda_i	\\
	\Rightarrow & \text{det.}[\Sigma \Sigma^*] = \Lambda_{+} \Lambda_{-}	\\
	\Rightarrow & 1=\Lambda_{+} \Lambda_{-}	\\
	\Rightarrow & \Lambda_{-}=\dfrac{1}{\Lambda_{+}}
\end{align*}
as $\Sigma$ and $\Sigma^*$ are both $U(3)$ matrices and complex conjugation of each other, $\Sigma \Sigma^*$ is a $SU(3)$ matrix. Note that, $Q$ being a $SU(3)$, its eigenvalues are uni-modular, which also implies, $\Lambda_{+}= e^{i \lambda}$ and $\Lambda_{-}=e^{-i \lambda}$ are complex conjugation of each other. Hence the eigenvalue equation corresponding to Eq. \ref{Q_and_Sigma_matrices} can be written as
\begin{equation}
	Q \psi_i = \{ e^{i \lambda},1,e^{-i \lambda}\} \psi_i.
	\label{Eigen_value_equation_SM}
\end{equation}

\section{Effective Hamiltonian}
\label{Effective_Hamiltonian}
We start by noting the fact that the matrix $\Sigma$ in Eq. \ref{Q_and_Sigma_matrices} is a $U(3)$ matrix. For the sake of clarity, let us write $Q$ as follows:
\begin{equation}
    \Sigma=
    \begin{bmatrix}
        r_{11}  &t_{12} &t_{13} \\
        t_{21}  &r_{22} &t_{23}  \\
        t_{31}  &t_{32} &r_{33}
    \end{bmatrix}
\end{equation}
where the elements $r_{ii}\rightarrow r_{ii} e^{i \phi_i}$ and $t_{ij}\rightarrow t_{ij} e^{i \phi_j}$ has implicit dependence on the superconducting phases. We define the effective Hamiltonian $\mathcal{H}$ in terms of this $\Sigma$ matrices:
\begin{equation}
    \mathcal{H}= \dfrac{1}{2 i} \left( \Sigma \Sigma^* -(\Sigma \Sigma^*)^{\dagger} \right)
\end{equation}
Now, $\mathcal{H}$ being Hermitian with $\text{Tr.}\mathcal{H}=0$, we can write the effective Hamiltonian in terms of 8 Gell-Mann matrices
\begin{equation}
    \mathcal{H}=\sum_{i=1}^{8} a_i \Tilde{\lambda}_i
\end{equation}
where $a_i \in \mathds{R}$ and $\text{Det.}\mathcal{H}=0$. Now we can take projection of $\mathcal{H}$ into each of these sub-spaces, say, $\mathcal{H}_1=a_1 \Tilde{\lambda}_1 + a_2 \Tilde{\lambda}_2 + a_3 \Tilde{\lambda}_3$, $\mathcal{H}_2= a_4 \Tilde{\lambda}_4 + a_5 \Tilde{\lambda}_5 + a_8 x$ and $\mathcal{H}_3 = a_6 \Tilde{\lambda}_6 + a_7 \Tilde{\lambda}_7 + a_8 y$. It is very easy to check with the choice of $x$ and $y$ that $\mathcal{H}_1+\mathcal{H}_2+\mathcal{H}_3=\mathcal{H}$. These projected Hamiltonians can describe individual topology in their own sub-spaces in a sense that whether in the parameter space of the independent superconducting phase differences, these Hamiltonians enclose their respective singular points.

As long the topology is concerned, the most crucial point in the parameter space is the point where $\mathcal{H}$ is singular i.e. all the eigenvalues of $\mathcal{H}$ are zero. A Hermitian matrix with all eigenvalues zero, is nothing but a null matrix, which means at the singular point all $\mathcal{H}_i=0$ or all the projections are identically zero.

Now, let, one of the projection, say $\mathcal{H}_1=0$ i.e. $a_1=a_2=a_3=0$. Then $\text{Det.}\mathcal{H}=0$ implies
\begin{equation}
   a_8 (a_4^2+a_5^2+a_6^2+a_7^2+2a_8^2)=0 
\end{equation}
which means either $a_4=a_5=a_6=a_7=a_8=0$ i.e. $\mathcal{H}_1=\mathcal{H}_2=\mathcal{H}_3=0$ or $a_8=0$ i.e. all the diagonal elements of $\mathcal{H}$ have to be zero. Now, the first case immediately means $\mathcal{H}=0$. We will show that in the second case for $a_8=0$, $\mathcal{H}=0$.

Let us write $\mathcal{H}$ explicitly in terms of $r_{ii}$ and $t_{ij}$.
\begin{equation}
    \mathcal{H}=\dfrac{1}{2i}
    \begin{bmatrix}
        \xi_{11}   &\xi_{12} &\xi_{13}  \\
        -\xi_{12}^* &\xi_{22}   &\xi_{23}   \\
        -\xi_{13}^* &-\xi_{23}^* &-(\xi_{11}+\xi_{22})
    \end{bmatrix}
    \label{H_xi_expression}
\end{equation}
where 
\begin{align*}
    \xi_{11}&=t_{12}t_{21}^*-t_{12}^* t_{21} +t_{13}t_{31}^* -t_{13}^*t_{31} \\
    \xi_{22}&= -t_{12} t_{21}^* + t_{12}^* t_{21} +t_{23} t_{32}^* - t_{23}^* t_{32}    \\
    \xi_{12}&= r_{11}(t_{12}^*-t_{21}^*)+r_{22}^* (t_{12}-t_{21})+t_{13} t_{32}^* - t_{23}^* t_{31}\\
    &= 2 r_{11}(t_{12}^*-t_{21}^*)+(t_{13}+t_{31})(t_{32}^*-t_{23}^*)   \\
    &=2 r_{22}^* (t_{12}-t_{21})+(t_{32}^*+t_{23}^*)(t_{13}-t_{31})   \\
    \xi_{13}&=r_{11}(t_{13}^*-t_{31}^*)+ r_{33}^* (t_{13}-t_{31}) +t_{12}t_{23}^* -t_{32}^* t_{21}  \\
    &= 2 r_{11} (t_{13}^*-t_{31}^*) + (t_{12}+t_{21})(t_{23}^*-t_{32}^*)    \\
    &= 2 r_{33}^* (t_{13}-t_{31})+(t_{23}^*+t_{32}^*)(t_{12}-t_{21})    \\
    \xi_{23}&=r_{22}(t_{23}^*-t_{32}^*)+ r_{33}^* (t_{23}-t_{32}) + t_{21} t_{13}^* - t_{31}^* t_{12}   \\
    &= 2 r_{22}(t_{23}^*-t_{32}^*)+(t_{21}+t_{12})(t_{13}^*-t_{31}^*)   \\
    &= 2 r_{33}^*(t_{23}-t_{32}) + (t_{13}^*+t_{31}^*)(t_{21}-t_{12}) 
\end{align*}
The projected Hamiltonians, in terms of $\xi_{ij}$, are
\begin{align*}
    \mathcal{H}_1 &=\dfrac{1}{2i}
    \begin{bmatrix}
        \dfrac{\xi_{11}-\xi_{22}}{2}    &\xi_{12}   &0   \\
        -\xi_{12}^* &-\dfrac{\xi_{11}-\xi_{22}}{2}  &0  \\
        0   &0  &0
    \end{bmatrix}   \\
    \mathcal{H}_2 &=\dfrac{1}{2i}
    \begin{bmatrix}
        \dfrac{\xi_{11}+\xi_{22}}{2}    &0    &\xi_{13}   \\
        0   &0  &0  \\
        -\xi_{13}^* &0 &-\dfrac{\xi_{11}+\xi_{22}}{2}
    \end{bmatrix}   \\
    \mathcal{H}_3 &=\dfrac{1}{2i}
    \begin{bmatrix}
        \dfrac{\xi_{11}+\xi_{22}}{2}    &\xi_{23}   \\
        -\xi_{23}^* &-\dfrac{\xi_{11}+\xi_{22}}{2}
    \end{bmatrix}
\end{align*}
Now if $\mathcal{H}_1=0$, then we have $\xi_{11}=\xi_{22}=\xi_{12}=0$. Now, $\xi_{12}=0$ implies,
\begin{equation}
    2 r_{11}=\dfrac{(t_{13}+t_{31})(t_{23}^*-t_{32}^*)}{(t_{12}^*-t_{21}^*)}, \hspace{20pt}
    2 r_{22}=\dfrac{(t_{32}+t_{23})(t_{31}^*-t_{13}^*)}{(t_{12}^*-t_{21}^*)}
\end{equation}
Using these, we can show that
\begin{equation}
    \xi_{13}=\dfrac{(t_{23}^*-t_{32}^*)}{(t_{12}^*-t_{21}^*)} (-\xi_{11})=0
\end{equation}
as $\xi_{11}=0$ and
\begin{equation}
    \xi_{23}= \dfrac{(t_{31}^*-t_{13}^*)}{(t_{12}^*-t_{21}^*)} (-\xi_{22}) =0
\end{equation}
as $\xi_{22}=0$ implying both $\mathcal{H}_2=0$ and $\mathcal{H}_3=0$. his proves that the singular point of $\mathcal{H}$ coincides simultaneously with the singular points of projected Hamiltonians $\mathcal{H}_1$, $\mathcal{H}_2$ and $\mathcal{H}_3$ or stating it physically, the gap closing in $\mathcal{H}$ and in its projected Hamiltonians $\mathcal{H}_1$, $\mathcal{H}_2$, $\mathcal{H}_3$ always occurs simultaneously.

\begin{figure*}[h]
    \centering
	\includegraphics[width=\columnwidth]{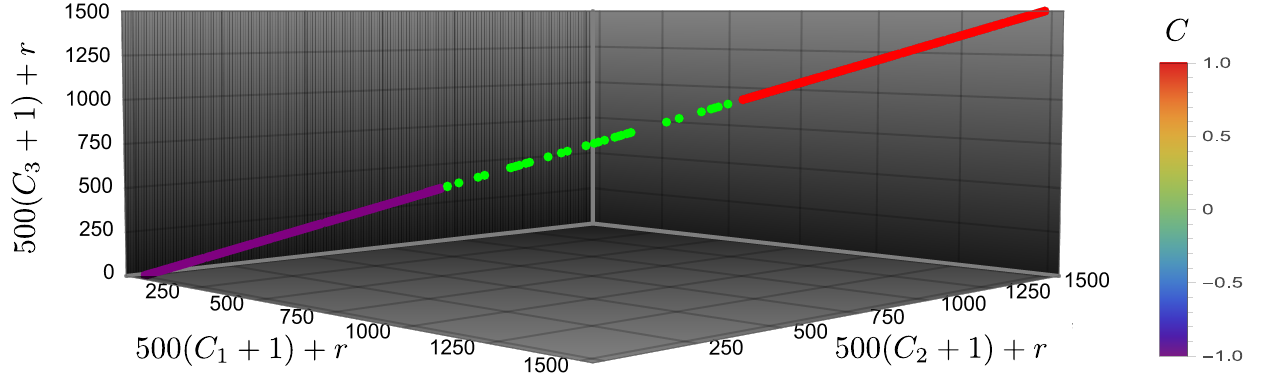}
        \caption{Chern number corresponding to wave functions $\psi_{-}$ and the gap in the novel spectrum $\Omega$. Torus formed by the image of Brillouin zone in the projected Hamiltonian $H_1$ are also shown at some critical points.}
	\label{Chern_500}
\end{figure*}

It is intriguing to check and compare the Chern numbers of $\mathcal{H}_1$, $\mathcal{H}_2$, $\mathcal{H}_3$ and $\mathcal{H}$. We have conducted a numerical simulation, as shown in FIG. \ref{Chern_500}, over 500 random $S$ matrices, chosen from Circular Unitary Matrix distribution and noted $500(C_i+1)+r$, $i \in \{1,2,3\}$, where $C_i$ is the Chern number of the negative band of $\mathcal{H}_i$ and $r$ represent the $r$th random unitary matrix. The points are color coded according to the Chern number of $\mathcal{H}$. We have found that the Chern numbers of these projected Hamiltonians and the original Hamiltonian always coincides.

\section{Quantum Hall based 3-JJ device}
\label{Device_fig}
\begin{figure*}[h]
    \centering
	\includegraphics[width=0.5\columnwidth]{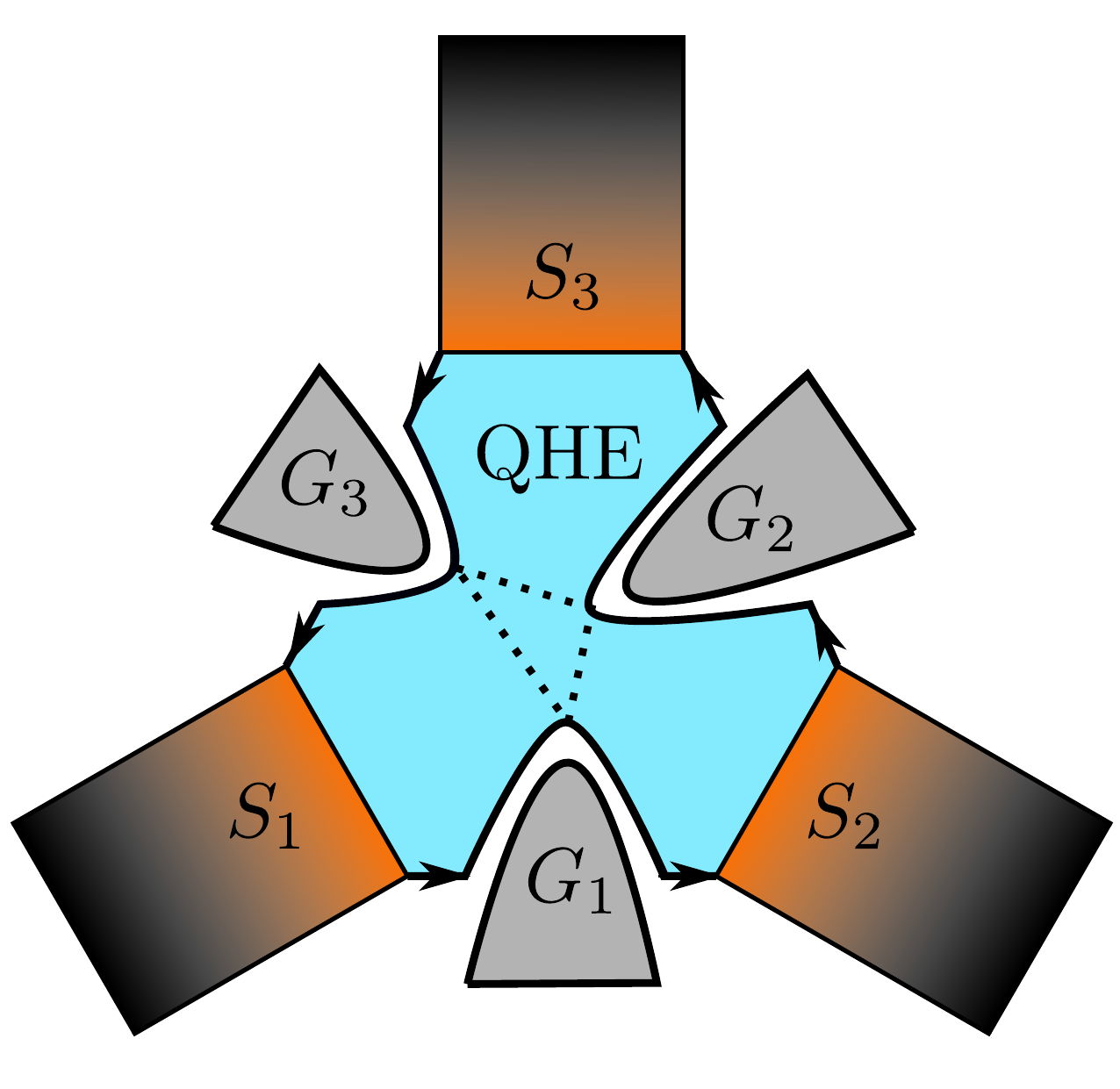}
        \caption{Topological 3-JJ in the presence of a quantum Hall edge (QHE) state at the junction region. Here $S_1$, $S_2$ and $S_3$ are the three superconductors. The scattering matrix at the junction can be controlled by tuning the voltage biases at the gates $G_1$, $G_2$ and $G_3$. Such device can be used to manipulate Majorana fermions at the junction as well as for storing the manipulated information safely by means of chirality after the manipulation. The dotted lines denote the tunnelling between different edges.}
	\label{QHE_QI_Store}
\end{figure*}

\end{document}